# A Comprehensive Study on Automated Testing with the Software Lifecycle

Hussein Mohammed Ali, Mahmood Yashar Hamza, Tarik Ahmed Rashid

**Abstract**— The software development lifecycle depends heavily on the testing process, which is an essential part of finding issues and reviewing the quality of software. Software testing can be done in two ways: manually and automatically. With an emphasis on its primary function within the software lifecycle, the relevance of testing in general, and the advantages that come with it, this article aims to give a thorough review of automated testing. Finding time- and cost-effective methods for software testing. The research examines how automated testing makes it easier to evaluate software quality, how it saves time as compared to manual testing, and how it differs from each of them in terms of benefits and drawbacks. The process of testing software applications is simplified, customized to certain testing situations, and can be successfully carried out by using automated testing tools.

**Index Terms**— Automated testing, Manual testing, Software testing, Software quality.

## 1 Introduction

One of the essential parts of the software development life cycle is testing the software. It's a set of tasks aimed at making sure that the software operates as it should, is easy to use, and does what it's supposed to do. Software testing is done to find bugs or mistakes [1]. The utilization of automation in software testing has significant importance due to its ability to improve operational efficiency by enabling faster test completion. Moreover, it ensures the accuracy of tests, thereby achieving a wider range of test coverage, particularly in complicated cases. Additionally, automation maintains uniformity in testing procedures, thereby mitigating the occurrence of human inaccuracies. Furthermore, it facilitates efficient regression testing and enables adjacent execution of tests. Ultimately, the implementation of automation in software testing results in the delivery of superior-quality software with a reduced number of flaws [2]. The main objective of automated software testing is to achieve certain objectives. This includes an exploration of the many testing tools commonly used in the industry, an understanding of successful approaches for developing automated test cases, and the mastery of the process involved in scripting and running these tests quickly [2]. Software development is a complicated task, and as the software gets larger, its expansion makes it more difficult to handle, and it gets harder to keep the quality together. Software testing provides an environment where you can work to improve quality. Because of quality concerns, numerous businesses spend almost half of their resources and time on software testing [3]. This article explores several different automated testing research articles regarding software requirements, software performance, and software testability. In the absence of automated testing, it becomes necessary to deal with human interaction, apart from the problem that we need more people and more time to do the testing, It is a well-established observation that human beings have a tendency to commit mistakes, often the testers get tired of doing the work repeatedly, and as a result, the utilization of the tests are more likely to have inaccuracies or failure. This paper displays automated testing tools that help us to do the testing with less human resources and make repeated tests done easily with fewer errors.

The upcoming sections of the paper are ordered as sections below. The problem statement of automated testing, in general, is briefly explained in section 2, the examination of the five projects and discussion of relevant articles works in section 3, the usage of automated testing in industries in section 4, the benefits of automated testing in section 5 and disadvantages in section 6, the tools and frameworks in section 7, the test cases of automated testing in section 8, and this paper's results and a discussion of relevant work are presented in section 9.

## 2 Problem Statement

Automated testing has advantages as well as drawbacks. Maintaining up-to-date testing following evolving software requires both time and expertise. One of the challenges encountered in the context of tests is the difficulty in effectively handling real-time changes while ensuring the proper management of the test data poses a complex task. The initial configuration process is complex, and the tests conducted may provide inaccurate outcomes, including both false positives and false negatives. Acquiring proficiency in testing tools may be challenging, and their ability to detect a typical flaw may be limited. The use of automation necessitates the involvement of individuals with specialized skills, while also potentially overlooking certain security concerns. Testing complex software requires a significant investment of time and effort.

## 3 Related Work

This section will conclude several software-related testing procedures, which have been applied in a variety of industrial processes with respected research articles as listed below.

### 3.1 Asset Management System for Distribution

In this research article, the system must track the delivered hardware locations, as well as the billing and the configurations that are done in the system. For the productivity of the system, test specialists should find issues that are related to the failure of the attempts that are made while employees using the system, to keep the system productive and maintainable. Their main goal was to create a system for best practices used for the testing and quality assurance of a web-based assets management system [4].

### 3.2 Java-based Application Platform

The primary objective of this research article is to establish a standard development platform for JAVA-based applications. Most of their concerns were centered on framework quality, testing procedures, auditing, and recording systems [4]. Following the analysis, a collection of reference applications was developed to evaluate the various components of the platform. Also, a few manuals were used for this purpose [5].

### 3.3 Point of Scale System for Life Insurance

Point-of-sale (POS) research was conducted for the finance department of a bank. The bank intended to provide a comprehensive selection of life insurance products. Instead of creating an entirely new sales institution, the existing branch offices could be utilized. Therefore, the POS system should permit finance personnel and bank employees to purchase insurance coverage for bank customers [6]. The foundation of requirements engineering is to utilize case studies and exploratory prototype testing. The POS system was implemented in a such way that, in the back-end of the system relational databases were used, CORBA as middleware, and Java as the primary programming language for both the business components of the middleware and as well as the client components [4].

### 3.4 Sales Assistance for Customized Industrial Facilities

The evaluation of this project, which intended to serve a significant worldwide industrial corporation, was like supporting sales efforts. This endeavor involved developing digital tools to assist their sales team. The primary objective was to swiftly create proposals for personalized industrial setups. This initiative followed a development approach similar to XP, fostering close collaboration with the client. This was crucial due to the challenge of managing changing requirements from diverse national sales divisions. The approach centered on leveraging automated tests using techniques like mock objects. Daily integration coupled with comprehensive automated test execution played a central role. Notably, it integrated the test architecture as a component of the overall system design. Remarkably, the automated test code accounted for about 25% of the complete codebase. This underscores the substantial impact of testing in ensuring effective system functionality [4].

### 3.5 Automation of System Tests for Control System

Within the context of System Test Automation for Control Systems, the project where a client grappled with extended-release cycles for their vital safety-critical control and information system. This application, designed as a distributed system, offered customization potential by configuration to suit individual customer requirements. The interconnected components communicated via a message bus, while the system remained compatible with diverse hardware platforms. Notably, the user interface could be configured for a distinct look and feel, enhancing user experience [4].

## 4 AUTOMATED TESTING IN INDUSTRY

As industrial automation systems increasingly rely on software, the significance of software quality has grown immensely. This paper addresses the imperative need for robust software quality assurance in automation projects. A fundamental aspect of this is the automated testing architecture proposed herein for software applications within industrial automation systems. The focus centers on testing the software of programmable logic controllers (PLCs) used in machinery, employing test automation frameworks that derive from software development practices. Practical insights stemming from software engineering best practices are synthesized to illustrate how these strategies can be transposed effectively to the domain of industrial automation. The amalgamation of these strategies materialized in an automation project and tests to validate the integrity of pivotal PLC components [7].

## 5 BENEFITS OF AUTOMATED TESTING

Test automation is an area of study in which there are essential approaches that must be done to accomplish it, however, it is known that there is also a gap between research work and the pros and cons of using the testing process, as shown in Fig. 1.

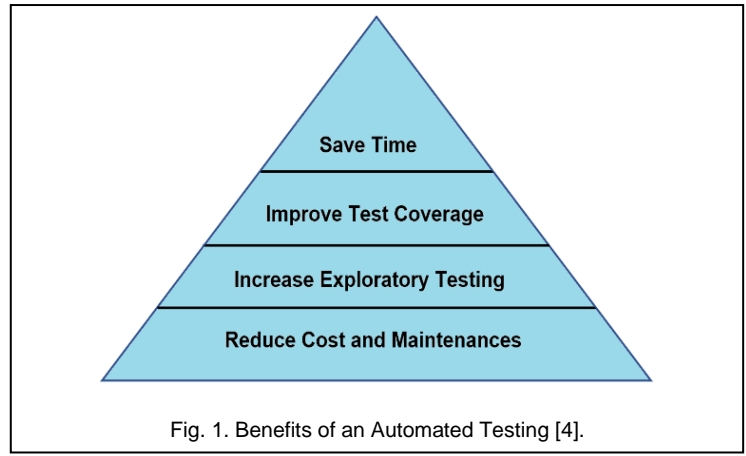

Fig. 1. Benefits of an Automated Testing [4].

Both professionals and researchers recognize the fact mentioned earlier. The aim of test automation is complete self-operation. Automated testing can be helpful in various ways, with numerous positive results suggesting that it can save money and address specific testing problems. However, even when automation has been beneficial, there are still many cases of failure, regrets, and negative feelings. Instances where previous attempts at automated testing have failed, despite significant investments and years of work in building the automation. When automation is introduced, work methods change, testing techniques shift, and even the testing process itself transforms. Automation can impact the products being tested and the processes of development and release, bringing both advantages and disadvantages that need careful evaluation [8]. Under S. Berner's paper, the benefits of automated testing include time savings for repetitive tasks, enhanced coverage for a wide range of scenarios, more room for exploratory testing to discover new errors, and cost reduction and easier maintenance through automation, which minimizes the need for human intervention and cuts down on resources and expenses for test case implementation. As we've discussed the advantages of automated testing, we'll now compare manual and automated testing in TABLE 1.

TABLE 1
COMPARISON BETWEEN AUTOMATED TESTING AND MANUAL TESTING

| AUTOMATED TESTING | MANUAL TESTING |
| --- | --- |
| It saves money. | It needs more money. |
| It takes less time. | Need more time. |
| It is reliable and increases confidence. | It is not as reliable as automated testing. |
| It is reusable and needs less human effort. | It needs more human effort. |
| Increases in fault and error detection. | Fault and error detection is not as high as automated testing. |
| Helps improve the quality of testing jobs and the efficiency of software. | Improving the quality is less compared to Automated testing. |
| Early time to market compared to Manual. | Late time to market compared to automate. |
| Since many testing tools may be utilized simultaneously, test coverage is increased. As a result, several test cases may be examined concurrently. A template for an automated test may be used repeatedly. | Often humans get tired of doing something repeatedly and they will take more time to do it. |

## 6   DISADVANTAGES OF AUTOMATED TESTING

In discussing the advantages of automated testing, it is crucial to acknowledge that it is not a solution and does present some challenges. Various tools have been developed that adapt to the diverse computer languages and testing methodologies. However, enterprises are faced with the task of selecting a single tool that can effectively address most of their testing requirements. Selecting the appropriate tool takes much effort, time, and planning for future endeavors. Proficiency in using the testing tool is a must. Several issues arise in this context, including the failure to achieve predetermined objectives and difficulties in maintaining the functionality of test scripts. The development of test automation requires a significant amount of time, as there are misconceptions among individuals about its implementation. Furthermore, the absence of expert personnel capable of effectively using test automation technologies poses a challenge, since expertise is necessary to compose automated test cases. The expenses associated with acquiring the testing tool and maintaining it for repeated approaches are rather expensive. [9] [10].

## 7 TOOLS AND FRAMEWORKS

Software testing automation tools can be used for many kinds of testing, such as unit testing tools, functional testing tools, code coverage tools, test management tools, and performance testing tools, as shown in Fig. 2. An engineering concept commonly integrated into automated testing is "Continuous Integration" (CI). Continuous Integration is a software development practice that involves regularly integrating code changes from multiple developers into a shared repository. Automated tests are a key component of CI, as they are executed automatically whenever new code is integrated. This practice helps ensure that new changes do not introduce defects or break existing functionality. By running automated tests as part of the integration process, developers receive rapid feedback on the health of the codebase, allowing issues to be identified and fixed early in the development cycle. Continuous Integration promotes collaboration, reduces integration problems, and contributes to maintaining a high-quality codebase in software engineering projects [11] [12].

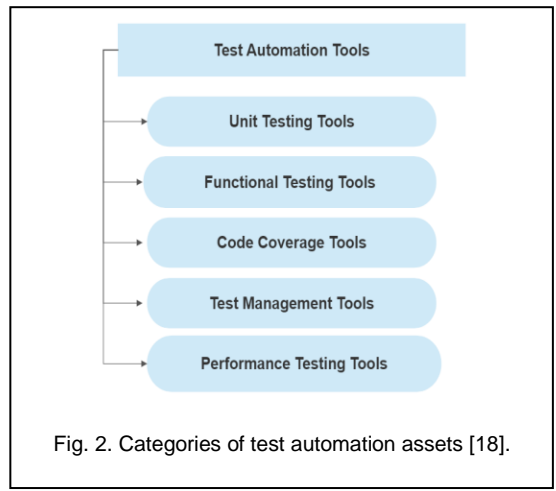

Fig. 2. Categories of test automation assets [18].

### 7.1 Unit Module Testing Tools

This type of testing tool delivers the validation of core code elements. These tools are essential in automating testing procedures, seamlessly integrated into development platforms like NetBeans. By employing automated testing, developers can execute test cases automatically using programming languages. By checking the code structure and adhering to best practices, unit testing tools play a pivotal role in verifying the precise operation of specific units or methods, thus ensuring their correct functioning, as well as going through each line code in the program [13]. Some of the unit testing frameworks are JUnit, NUnit, JMockit, and PHPUnit which are represented in

TABLE 2.

### 7.2 Function Testing Tools

This kind of testing is conducted to ensure that the software functions in alignment with the requirements of the users. In the course of doing functional testing, we use the functional testing tools at our disposal. Functional testing tools are used to check the performance of functions by supplying them with input and afterward comparing the actual output of these functions with the expected outcome for the specific test case. Functional testing tools are used to assess the extent to which a software system adheres to the specified requirements. The user's text does not contain any information to rewrite. Several functional testing tools often used in software testing include Selenium, HP QuickTest Professional, TestComplete, Ranorex, Watir, Tricentis Tosca Testsuite, and Test Studio, among others [14], which are represented in

TABLE 2.

TABLE 2
UNIT AND FUNCTION TESTING TOOLS

| UNIT MODULE TESTING TOOLS | FUNCTION TESTING TOOLS |
|---|---|
| **JUnit:** is widely used in the Java environments. It gives the ability to the developers to establish and run replicable tests, guaranteeing the accuracy of discrete code fragments such as methods or functions. With annotations and assertions, JUnit aids in defining and validating test cases, thereby enhancing code quality, and detecting issues early on.<br>**NUnit:** employs to the .NET domain, compatible with programming languages like C#. Like JUnit, NUnit enables the creation and execution of tests for distinct code components. NUnit boasts features like parameterized tests, setup/teardown methods, and test runners, bolstering the effectiveness of the testing process.<br>**JMockit:** Stands out as a library for unit testing, with a particular focus on mocking techniques. Mocking replicates the behavior of external dependencies, isolating the code under examination from external influences. JMockit simplifies this by crafting mock implementations for dependencies, enhancing precision in intricate testing scenarios.<br>**PHPUnit:** Caters to PHP, dedicated to verifying the reliability of PHP codebases. It equips developers with tools for composing, executing, and managing tests. With test fixtures, assertions, and test suites, PHPUnit ensures accurate functioning and assists in ongoing system maintenance. This practice is prevalent in PHP development for bug identification, correctness assurance, and system maintenance. | **Selenium:** Selenium is an open-source automation framework primarily used for web application testing. It provides a suite of tools to automate browser interactions and validate web elements. Selenium WebDriver is widely used for browser automation, while Selenium IDE allows users to create and run tests through a browser extension.<br>**HP QuickTest:** HP QuickTest Professional, now known as Micro Focus Unified Functional Testing (UFT), is a commercial tool for automated functional testing. It supports a wide range of technologies, including web, mobile, desktop, and API testing. It offers features like keyword-driven testing and record-and-playback.<br>**TestComplete:** TestComplete is a commercial automation tool by SmartBear that supports various application types, including web, desktop, mobile, and API testing. It offers a script-free testing approach and supports multiple scripting languages.<br>**Ranorex:** Ranorex is an automation tool designed for testing desktop, web, and mobile applications. It provides a user-friendly interface for creating automated tests without extensive coding knowledge. Ranorex supports multiple programming languages. |

The data sources used by these technologies in the context of automated testing often include the interaction with diverse data sets to verify the behavior of software. Customizing tools and data sources to align with the requirements and contextual factors of testing scenarios has significant importance. Common sources of data often used in academic research include:

- **Test Data Files:** These files include test input data in various formats such as CSV, Excel, JSON, or XML.
- **Databases:** Automated tests can interact with databases to verify data integrity and system behavior.
- **API Responses:** Testing APIs involves sending requests and verifying responses, often in JSON or XML formats.
- **UI Elements:** Data-driven tests can use data sources to populate forms or simulate user interactions.
- **External Files:** Applications that process files can be tested by providing sample files for validation.

## 7.3 Code Coverage Tools

These automated tests are used to identify which sections of the software code have been tested successfully by the various testing tools. It tracks the number of lines, statements, or blocks of code that have been run through automated test suites for verification. It is an important measure to use to understand the quality of the Quality Assurance (QA) efforts that have been undertaken.

## 7.4 Software for Managing Tests

These technologies are used to automate test plans (such as developing test cases, test plans, test strategies, test results, test reports, etc.), and they assist teams in managing projects more efficiently by providing a placeholder for test activities that can be searched and maintained. The many different test management solutions each come with their own set of features and methods for managing the testing process. On the other hand, they often provide the opportunity to simplify the testing process, make it possible to have instant access to data analysis and facilitate communication across various project groups [15].

## 7.5 Performance Testing Tools

During the performance testing procedures, performance testing techniques are utilized. The purpose of these tests is to evaluate the responsiveness and stability of the program across a wide variety of scenarios and demand levels. In addition, it can be used to investigate, measure, validate, or verify other aspects of software quality, including scalability, dependability, and resource utilization. These methods help in the process of evaluating software or component performance in terms of resource use, throughput, and stimulus-response time following a set of required performance standards. In addition to endurance and spike testing, additional performance testing categories include soak, breakpoint, configuration, isolation, and Internet testing [16].

## 8  Test Cases on Automated Testing

Automation testing has limitations, not every test is automated. It is essential to determine which cases should be automated. Its benefits depend on the number of times the test can be performed [17]. Choosing the appropriate test cases is crucial because it affects the development budget and time. The suitable and unsuitable test cases are displayed in TABLE 3.

TABLE 3
SUTABILE AND NON-SUITABLE TEST CASES FOR AUTOMATED TESTING.

| SUITABLE TEST CASES | NON-SUITABLE TEST CASES |
| --- | --- |
| Tests that need to be run for every part of the system. | Test cases that are one-time done. |
| Tests that use multiple data values for the same action. | Automation testing is not used for usability. |
| Similar tests need to be executed using different platforms. | Tests that do not have predictable results. |
| Regression testing. | Test cases that are not executed manually at least once should not be done by automation testing. |
| Cases of mission-critical testing. | Systems that its requirements are frequently changing. |
| Test cases that must be executed repeatedly. | Execution of test cases on an ad hoc basis. |
| Time-consuming test cases. | |

## 9  Research on the Importance of Automated Testing

Automated testing has been a trending topic for researchers due to the increasing amount of software in both industry and academic fields. According to the Institute of Electrical and Electronics Engineers IEEE exploration database the amount of research regarding automation testing has been dramatically increasing throughout the years, as illustrated in Fig. 3, the amount of research on automated testing reached its peak point in the year 2022. As our research shows, the automated testing field has been a hot topic since the year 2000 and has been increasing gradually throughout the years, and even for now, the published article and papers count has nearly reached five thousand. The reason behind this is that the technology is increasing, and the skills required to test software are becoming much more difficult, as well as The increasing software, programming language types, and manual testing are less wanted by the companies, due to the workload and complex adaptation of the employees to the work.

## 10 CONCLUSION

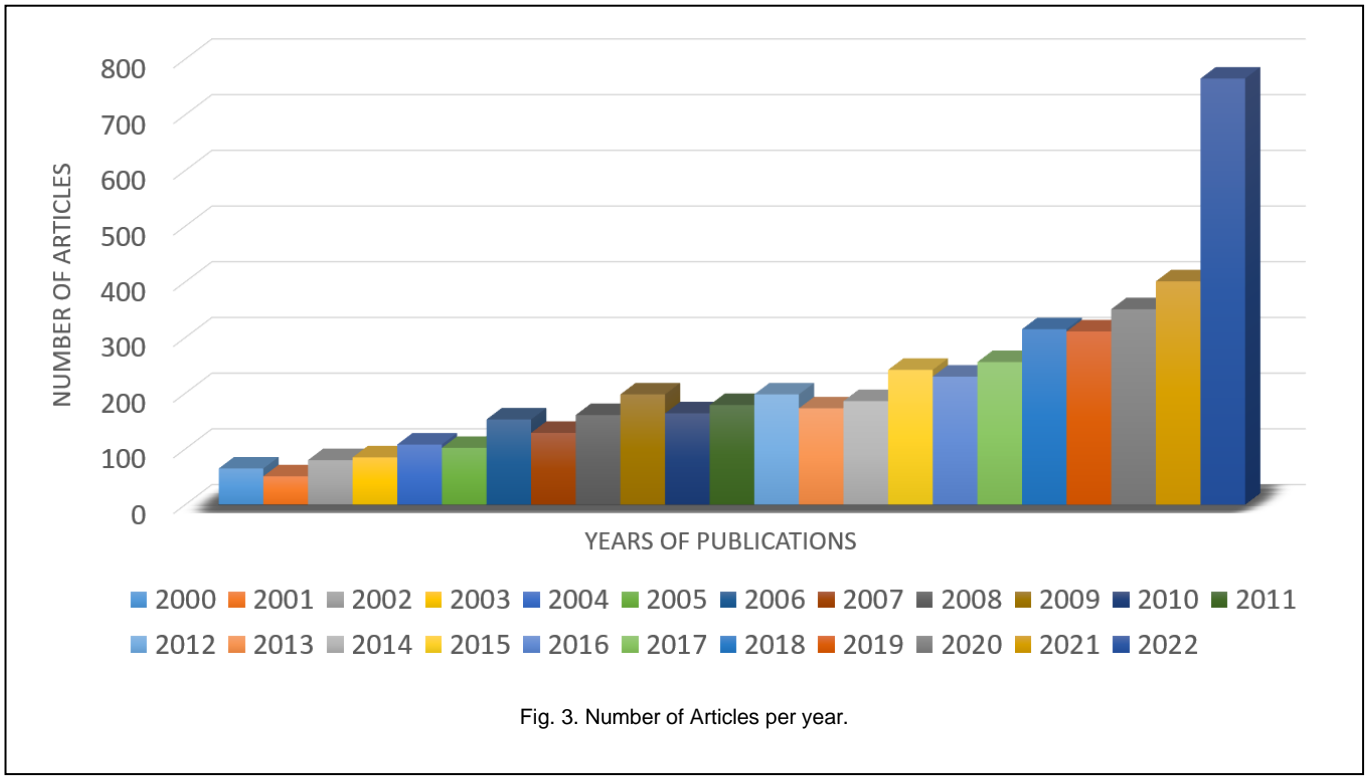

Fig. 3. Number of Articles per year.

This paper presented the essential parts of Automated Testing, in which the study contains a brief explanation of the topic, an overview of several relevant projects, with examples of how to improve their work, a solution to the problems they have faced during the implementations, as well as the benefits and drawbacks of the techniques of Automated Testing. Also, specific tools for implementing the tools according to the type, and the test cases for suitableness of the software. This paper aimed to conclude the main concepts of automation testing as stated in the sections above, automation testing is the core principle for checking the quality of the software components. Also, test management was stated, with some industrial testing examples. This paper has successfully provided foundational insights into Automated Testing, addressing these potential limitations and pursuing future research opportunities will contribute to a more comprehensive understanding of automated testing's evolving landscape and its potential to revolutionize software quality assurance.

For future directions of this paper, which are listed below:
1. **Dynamic Test Orchestration:** Research could focus on creating algorithms that dynamically select and adjust test cases based on real-time changes in software, optimizing testing resources and coverage.
2. **Ethical and Social Implications:** Exploring biases in test data, considering the role of human testers alongside automation, and ensuring fairness and transparency in automated testing practices.
3. **Real-world Monitoring:** Investigating continuous validation and monitoring of software in live environments to proactively identify performance issues and bugs, improving overall system reliability.

Findings in these areas could contribute to the development of more intelligent and adaptable testing systems, a deeper understanding of the societal impact of automation, and approaches that bridge the gap between development and real-world application.

————————————————

- *Author name: Hussein Mohammed Ali is currently working as a research assistant in the department of Computer Engineering at Tishk International University, Iraq. E-mail: hussein.mohammed@tiu.edu.iq*

- *Co-Author name: Mahmood Yashar Hamza is currently working as a research assistant in the department of Computer Engineering at Tishk International University, Iraq. E-mail: mahmood.yashar@tiu.edu.iq*
- *Co-Author name: Tarik Ahmed Rashid is currently serving as dean in Computer Science and Engineering at University of Kurdistan Hewler, Iraq - E-mail: tarik.ahmed@ukh.edu.krd*